\documentclass[fleqn,usenatbib]{mnras}

\usepackage{newtxtext,newtxmath}
\usepackage[T1]{fontenc}
\usepackage{ae,aecompl}
\usepackage{url}
\usepackage{graphicx}
\usepackage{xspace}
\usepackage{amsfonts}
\usepackage{pifont}
\usepackage{textcomp}
\usepackage{float}
\usepackage{subfig}
\usepackage{multirow}
\usepackage{adjustbox}
\usepackage{textgreek}
\usepackage{booktabs}

\newcommand{\kms}{\mbox{$\mathrm{km~s^{-1}}$}}
\newcommand{\Ha}{H\textalpha}

\newcommand{\hel}[2] {He\,{\sc #1}~\textlambda #2}
\newcommand{\Line}[3]{#1\,{\sc #2}\;\textlambda#3}

\title{ASASSN--14dx: A cataclysmic variable harbouring a massive pulsating white dwarf}

\author[P. Hakala et al.]
{Pasi Hakala$^{1}$\thanks{E-mail: pahakala@utu.fi},
Ingrid Pelisoli$^{2}$,
Boris T. G\"ansicke$^{2}$,
Pablo Rodríguez-Gil$^{3,4}$, 
Thomas R. Marsh$^{2}$,
\newauthor
Elm\'e Breedt$^{5}$,
John R. Thorstensen$^{6}$,
Anna F. Pala$^{7}$
\\
$^{1}$Finnish Centre for Astronomy with ESO (FINCA), Quantum, University of Turku, FI-20014 Finland\\
$^{2}$Department of Physics, University of Warwick, Coventry CV4 7AL, UK\\
$^{3}$Instituto de Astrofísica de Canarias, E-38205 La Laguna, Tenerife, Spain\\
$^{4}$Departamento de Astrofísica, Universidad de La Laguna, E-38206 La Laguna, Tenerife, Spain\\
$^{5}$Institute of Astronomy, University of Cambridge, Madingley Road, Cambridge CB3 0HA, UK\\
$^{6}$Department of Physics and Astronomy, Dartmouth College, Hanover NH 03755, USA\\
$^{7}$European Southern Observatory, Karl-Schwarzschild-Strasse 2, Garching, 85748, Germany.\\
}

\begin{document}
\label{firstpage}
\pagerange{\pageref{firstpage}--\pageref{lastpage}}

\outer\def\gtae {$\buildrel {\lower3pt\hbox{$>$}} \over 
{\lower2pt\hbox{$\sim$}} $}
\outer\def\ltae {$\buildrel {\lower3pt\hbox{$<$}} \over 
{\lower2pt\hbox{$\sim$}} $}
\newcommand{\Msun}{$\mathrm{M}_{\odot}$}
\newcommand{\lsun}{$L_{\odot}$}
\newcommand{\Rsun}{$R_{\odot}$}
\newcommand{\solar}{${\odot}$}
\newcommand{\kep}{\sl Kepler}
\newcommand{\ktwo}{\sl K2}
\newcommand{\tess}{\sl TESS}
\newcommand{\swift}{\it Swift}
\newcommand{\Porb}{P_{\rm orb}}
\newcommand{\nuorb}{\nu_{\rm orb}}
\newcommand{\eplus}{\epsilon_+}
\newcommand{\eminus}{\epsilon_-}
\newcommand{\cd}{{\rm\ c\ d^{-1}}}
\newcommand{\MdotL}{\dot M_{\rm L1}}
\newcommand{\Mdot}{$\dot M$}
\newcommand{\Mdsolar}{\dot{\mathrm{M}_{\odot}} yr$^{-1}$}
\newcommand{\Ldisk}{L_{\rm disk}}
\newcommand{\src}{KIC 9202990}
\newcommand{\ergscm} {erg s$^{-1}$ cm$^{-2}$}
\newcommand{\rchi}{$\chi^{2}_{\nu}$}
\newcommand{\chisq}{$\chi^{2}$}
\newcommand{\pcmsq} {cm$^{-2}$}

\maketitle

\begin{abstract}
We present the results of our study of ASASSN--14dx, a previously known but poorly characterised cataclysmic variable (CV). The source was
observed as part of an ongoing high-time-resolution photometric survey of CVs, which revealed that, in addition to the known 82.8-min orbital period, it also exhibits other transient periods, the strongest of which around 4 and 14 min. Here, we report our findings resulting from a multifaceted follow-up programme consisting of optical spectroscopy, spectropolarimetry, imaging polarimetry, and multicolour fast photometry.
We find that the source displays complex optical variability, which is best explained by the presence 
of a massive white dwarf exhibiting non-radial pulsations.
An intermediate polar-like scenario involving a spinning magnetic white dwarf can be ruled out based on the detected changes in the observed periods. Based on our optical spectroscopy, we can constrain the mass and effective temperature of the white dwarf to be $\sim$1.1\;$\mathrm{M}_{\odot}$ and 16\,100\;K, respectively. The overall intrinsic flux level of the source is unusually high, suggesting that there remains significant residual emission from the accretion disc and/or the white dwarf even ten years after the 2014 outburst. Finally, we cannot detect any spectroscopic signatures from the donor star, making ASASSN--14dx a possible period bouncer system evolving towards a longer orbital period.
\end{abstract}

\begin{keywords}
Physical data and processes: accretion, magnetic accretion -- stars: binaries close -- stars: white dwarfs -- stars: dwarf novae
\end{keywords}

\section{Introduction}

Cataclysmic Variables (CV) are semi-detached binaries consisting of a white dwarf (the "primary") accreting from a Roche lobe-filling main sequence star (the "secondary" or "donor").  The secondary transfers mass via Lagrangian L$_{1}$ point overflow. This is then accreted by the primary white dwarf, typically through an accretion disc \citep[see][for a review]{1995cvs..book.....W}. In case of a significantly magnetic ($>$\,2--5\;MG) white dwarf, its magnetic field prevents the formation of an inner accretion disc (in intermediate-polar CVs) or any accretion disc (in polar CVs), causing the accretion flow to be funnelled along the magnetic field lines of the white dwarf.
    
The accretion disc is usually present even in CVs that reside in a quiescent state (i.e. they are not undergoing an episode of enhanced accretion, such as a dwarf nova outburst, for instance).
Thus, the optical spectrum consists of three components: the light from the donor star, the white dwarf, and the emission from the accretion disc. 

\begin{figure*}
  \begin{center}
\hspace{0mm}  
\vspace{0mm}
  \includegraphics[width=0.95\textwidth,angle=0]{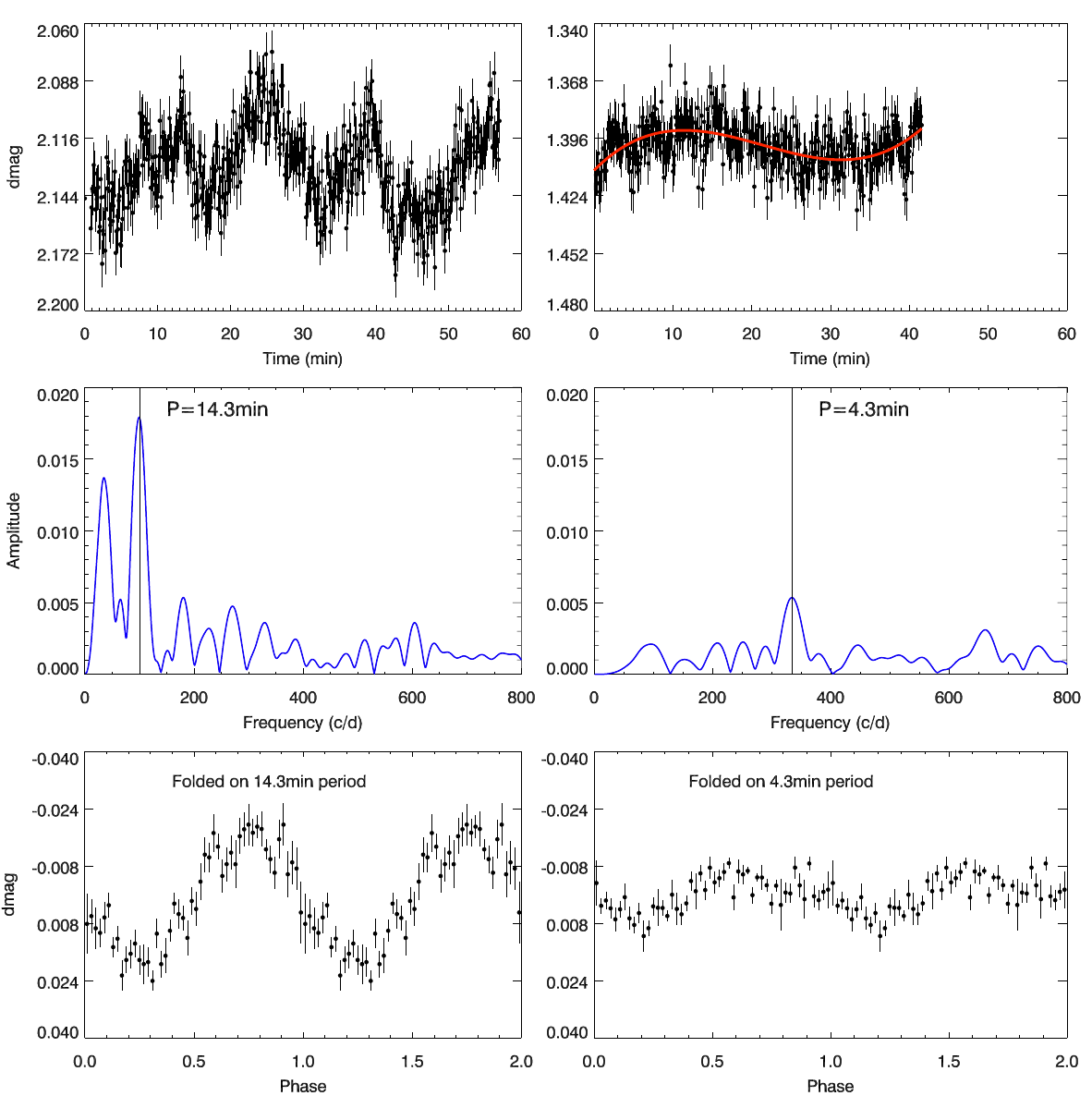}    
\vspace{0mm}
  \caption{The two 2021 NOT photometry sets in the sky contrast filter (26 Aug 2021 top-left and 31 Aug 2021, top-right), their Lomb-Scargle amplitude spectra (middle) and the data phase-folded and binned on the 14.3- and 4.3-min periods, respectively (bottom). A third-order polynomial (shown in red) was removed from the top-right light curve prior to the period analysis. We note that the 4.3-min period obtained from the 40-min NOT time-series differs slightly from the 4.1-min period determined from the 5-hour 2022 ULTRACAM observation. Given the brevity of the NOT run, the difference is not likely to be significant though.}
    \label{figurenotphot}
    \end{center}
\end{figure*}

CVs start their lives with orbital periods typically of the order of a few hours and gradually evolve towards shorter orbital periods as magnetic braking, and gravitational waves, remove angular momentum from the system. The evolution towards shorter orbital periods continues until mass loss causes the companion to become degenerate and stop responding to mass loss with a decrease in radius. At this point, the system reaches a period minimum (around 80\;min) and the orbit will start increasing \citep{2009MNRAS.397.2170G}. 

The orbital period can often be detected in CV light curves due to, for example, double humped light curve from an accretion disc in case of relatively high inclination system, where the accretion disc can partially eclipse (and be partially eclipsed by) the secondary star leading to two minima (at phases 0.0
and 0.5). Occasionally it is possible to detect the ellipsoidal modulation by the secondary star,
if it contributes enough to the overall flux. Sometimes the single humped modulation from the hot spot in the accretion disc, where the ballistic accretion stream hits the disc outer edge, can be detected. The modulation due to the hot spot could occasionally also be double humped, if the accretion disc is optically thin, but the hot spot is optically thick and elongated in shape \citep{2000MNRAS.318..429S}. Furthermore, the spin of the white dwarf can occasionally also be a component in the light curve, particularly for intermediate polars, where the funnelled accretion flow can create a bright spot on the white dwarf that rotates in and out of view \citep[see][for an early review]{1994PASP..106..209P}. Also, spiral shocks in the accretion disc can lead to the formation of hot spots on the white dwarf surface even without any magnetic field involvement \citep{2019MNRAS.483.1080P}. Tracking changes in the observed spin period \citep[see for instance][]{2020ApJ...897...70P,2021MNRAS.507.6132P,2024MNRAS.531L..82P} is a common pathway for improving our understanding of the complex angular momentum exchange mechanisms at play in CV evolution. An accreting white dwarf is spun up by the accreting matter \citep{1985A&A...148..207R,1991MNRAS.251P..30K}. On the other hand, the strong enough white dwarf magnetic field tends to synchronise the compact object spin to the orbital period \citep{1983MNRAS.205.1031C}. The interplay of accretion vs. magnetic torque has a strong bearing also on our understanding of white dwarf magnetic field generation and the evolution of different types of CVs \citep{2021NatAs...5..648S}. In addition to the white dwarf spin period, short-period (typically on the order of minutes) modulation in the white dwarf flux can also be caused by non-radial g-mode pulsations \citep{2021FrASS...8..184S}. These are excited by cyclic partial ionisation of the hydrogen and/or helium in the atmosphere of the white dwarf \citep{1979ApJ...229..203M}. 

\begin{table}
  \centering
  \begin{tabular}{lll}
    \hline
    \hline
    SDSS u & 16.417 & $\pm$ 0.007\;mag\\
    SDSS g & 16.256 & $\pm$ 0.004\;mag\\
    SDSS r & 16.376 & $\pm$ 0.005\;mag\\
    SDSS i & 16.563 & $\pm$ 0.005\;mag\\
    SDSS z & 16.648 & $\pm$ 0.009\;mag\\
    \hline
    Mean $G$ & 15.20 & $\pm$ 0.02\;mag\\
    Mean $(BP-RP)$ & 0.2 & $\pm$ 0.1\;mag\\
    $M_G$ & 10.64 & $\pm 0.02$\;mag\\
 \hline
 \hline
  \end{tabular}
  \caption{\textit{SDSS DR16} \citep{2020ApJS..249....3A} and \textit{Gaia} photometry of ASASSN--14dx}
  \label{table_Gaia}
\end{table}   

\begin{table*}
  \begin{tabular}{lccccccc}
    \hline
Telescope/instrument & Date & Duration (min) & Filter/grism & Observing mode \\
    \hline
NOT(2.56m)/ALFOSC & 26 Aug 2021 & 57 & sky contrast &  fast photometry  \\ 
NOT(2.56m)/ALFOSC & 31 Aug 2021 & 42 & sky contrast &  fast photometry  \\
NTT(3.5m)/ULTRACAM & 22 Aug 2022 & 287 & $u^\prime$$g^\prime$$r^\prime$ & fast photometry  \\
NOT(2.56m)/ALFOSC & 28 Nov 2022 & 85 & sky contrast & imaging circ. polarimetry \\
McGraw-Hill(1.3m)/Templeton & 14 Dec 2022 & 393 & GG420 & fast photometry \\
NOT(2.56m)/ALFOSC & 21 Dec 2022 & 125 & sky contrast & imaging circ. polarimetry \\
NOT(2.56m)/ALFOSC & 21 Dec 2022 & 2x5 & Grism \#4 & circ. spectropolarimetry \\
SAAO(1.9m)/SpUpNIC & 10 Feb 2023 & 4x10 & Grating \#6 & spectroscopy \\
NTT(3.5m)/ULTRACAM & 4 Sep 2023 & 300 & $u^\prime$$g^\prime$$r^\prime$ & fast photometry \\
VLT(8.2m)/X-Shooter & 18 Oct 2023 & 2x5 & & spectroscopy \\
NOT(2.56m)/ALFOSC & 15 Sep 2024 & 123 & Grism \#18 & phase resolved spectroscopy \\
\hline
  \end{tabular}
  \caption{The observing log in chronological order.}
  \label{table_phot_log}
\end{table*}

Another commonly observed type of variability in CVs is the sudden increase in brightness during outbursts \citep[likely caused by disc instability][]{2001A&A...370..488O}), which leads to their detection as transient sources. This was the case for ASASSN--14dx, which was detected by the All-Sky Automated Survey for Supernovae \citep[ASAS--SN,][]{2014ApJ...788...48S} on 25 June 2014, with an outburst magnitude of V=13.95 \citep{2019PASJ...71...22I}. A blue spectrum displaying broad
Balmer absorption lines with emission cores was reported by \citet{2014ATel.6624....1K}, while \citet{2016AJ....152..226T} discovered radial velocity modulation in the H$\alpha$ emission line cores. These emission features, presumably following the orbital motion of the white dwarf, showed a radial velocity amplitude ($K_1$) of 41\;\kms\ \citep{2016AJ....152..226T}. The source has an orbital period of 0.0575\;d (82.8\;min) \citep{2016AJ....152..226T}, and its \textit{Gaia} distance is 81.4 $\pm$ 0.2\;pc, making it one of the closest CVs \citep{2019PASJ...71...22I}. The \textit{Gaia} magnitudes, listed in Table~\ref{table_Gaia}, place the source about 1.7--1.8\;mag above the \textit{Gaia} white dwarf cooling sequence for $(BP-RP) = 0.2$, indicating substantial additional emission---likely from an accretion disc, as the secondary star is expected to be faint at the given orbital period due to its very low mass and late spectral type.

The blue optical spectrum displays very broad absorption features from a white dwarf, as well as strong double-peaked Balmer lines in emission \citep{2016AJ....152..226T}. The former suggests that ASASSN--14dx might contain a high-mass white dwarf. CVs with massive white dwarfs are intriguing as potential candidates for type Ia supernova (SNIa) progenitors \citep{2012JASS...29..163K}. 
\citet{2019PASJ...71...22I} proposed that the source belongs to the WZ\;Sge subclass of dwarf novae, based on the ASAS--SN discovery outburst properties, along with the 82.8-min orbital period. The WZ\;Sge class of CVs typically exhibits large outburst amplitudes, coupled with recurrence times of decades, and short orbital periods of 80--90\;min. For further details, see \cite{2015PASJ...67..108K}, and references therein. Here, we present the results of our time-series photometry, circular polarimetry, and optical spectroscopy of ASASSN--14dx, which we use to derive parameters for both the binary system and the white dwarf itself.

\begin{figure}
  \begin{center}
\hspace{0mm}  
\vspace{0mm}
  \includegraphics[width=0.47\textwidth]{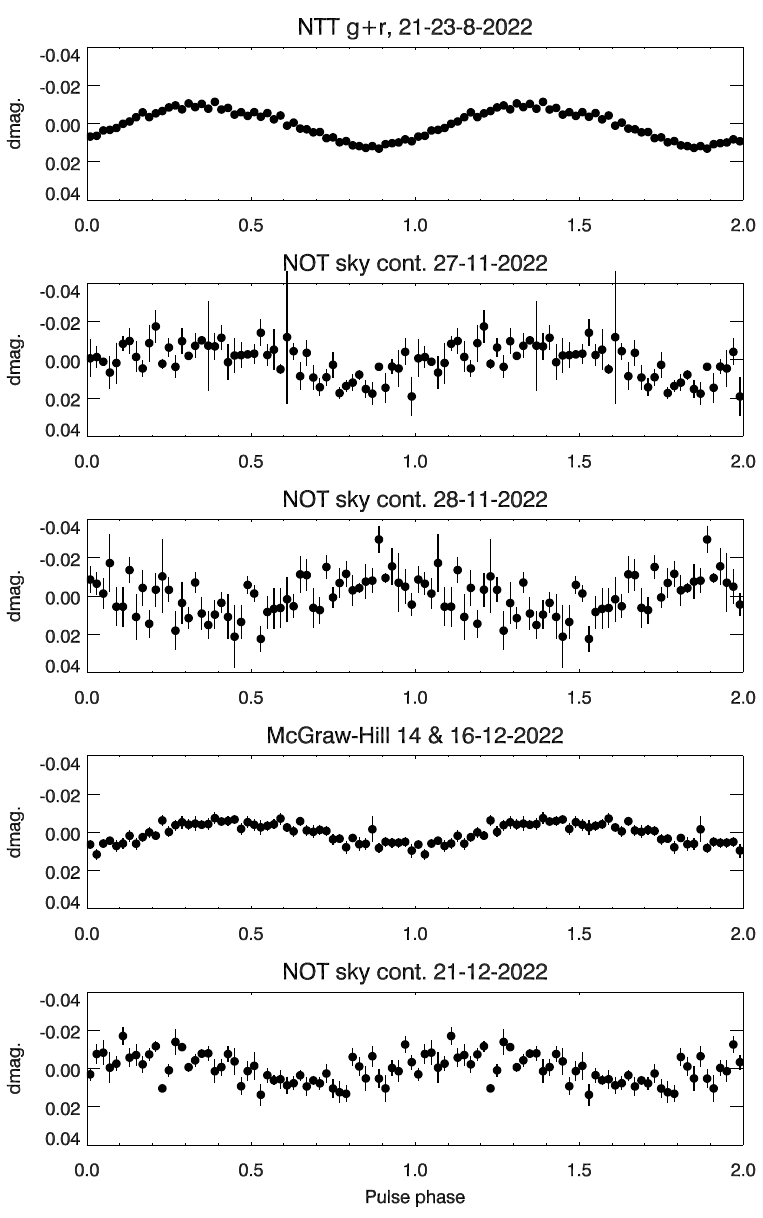}    
\vspace{0mm}
  \caption{The August, November and December 2022 photometry sets, phase-folded on the 4-min period. Note the phase reversal between the 27 and 28 November light curves.}
    \label{figurephotom}
    \end{center}
\end{figure}

\begin{figure*}
  \begin{center}
\hspace{-7mm}  
\vspace{0mm}
  \includegraphics[width=1.02\textwidth]{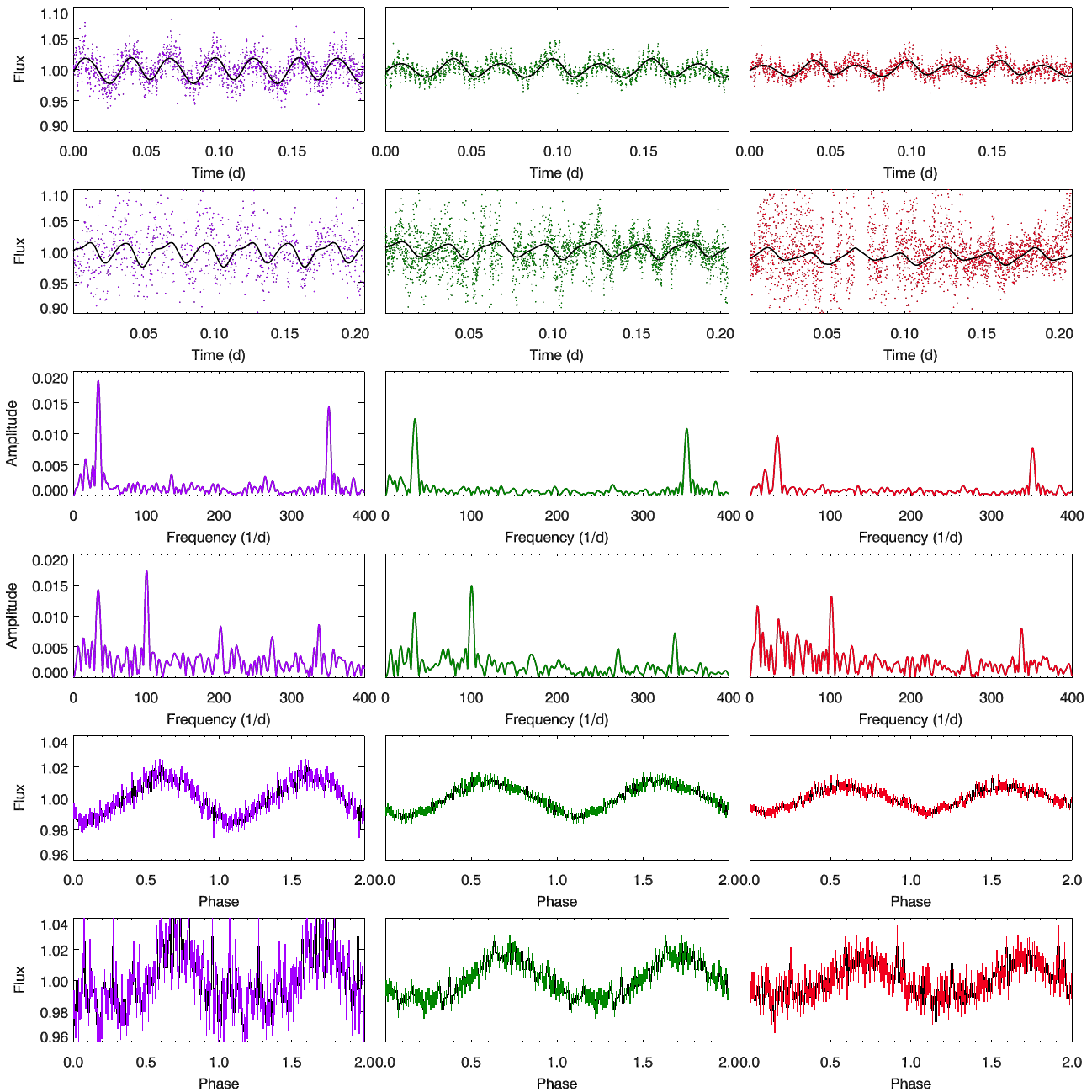}    
\vspace{0mm}
  \caption{The $u^\prime$, $g^\prime$, and $r^\prime$ photometry (columns from left to right) obtained with NTT/ULTRACAM in August 2022 (top row) and September 2023 (second row), along with the best-fit orbital modulation solutions (fourth-order Fourier series). The corresponding amplitude spectra are shown in rows 3 and 4, while the orbital modulation-subtracted light curves, folded on the 351 cycle\;d$^{-1}$ (August 2022) and 100 cycle\;d$^{-1}$ (September 2023) frequencies, are shown in the rows 5 and 6, respectively. Both the 4.1-min (351 cycle\;d$^{-1}$) and 14.3-min (100 cycle\;d$^{-1}$) modulations increase in amplitude towards the blue.}
    \label{figurentt}
    \end{center}
\end{figure*}

\section{Observations}
\subsection{Photometry}
Optical photometry was conducted at three different sites: the Nordic Optical Telescope (NOT) with ALFOSC on La Palma, the New Technology Telescope (NTT) with ULTRACAM at La Silla, and the 1.3-m McGraw-Hill telescope with the Templeton CCD detector at Kitt Peak. All data were bias-subtracted and flat-fielded in the standard manner for these setups.

ASASSN--14dx was initially observed as part of a campaign to search for fast-spinning white dwarfs in CVs (in prep.). These observations involved 60-min high time-resolution photometric time series
and were conducted at both the NOT and the NTT. The source was observed at several epochs following the detection of periodicity
in the first NOT data set (see Table~\ref{table_phot_log} for the observing log). 

The initial NOT photometry on 26 Aug 2021 revealed a 14-min period; however, the detection was tentative, as the 56-min time series covered only four cycles (see Fig.~\ref{figurenotphot}, left panels). A second NOT observation, taken on 31st Aug 2021, truncated to just 42 min in length, showed no indication of the 14\;min period. Instead, a weak signal at 4\;min was detected (Fig.~\ref{figurenotphot}, right panels). ASASSN--14dx was observed again with NTT/ULTRACAM in August 2022 for several hours. These observations clearly revealed the 4-min (351\;cycle\;d$^{-1}$) period in all $u^\prime$, $g^\prime$, and $r^\prime$ filters (Fig.~\ref{figurentt}). Subsequently, another set of NTT observations, using an identical instrument setup, was conducted a year later (in September 2023; see also Fig.~\ref{figurentt}). This time, only the 14-min (100\;cycle\;d$^{-1}$) period was clearly detected. Additional photometry was obtained with the 1.3-m McGraw-Hill telescope in December 2022. These observations also revealed only the 4-min period (Fig.~\ref{figurephotom}). 

\begin{figure*}
  \begin{center}
  \vspace{0mm}   
  \hspace{0mm}  
  \includegraphics[width=0.45\textwidth,angle=0]{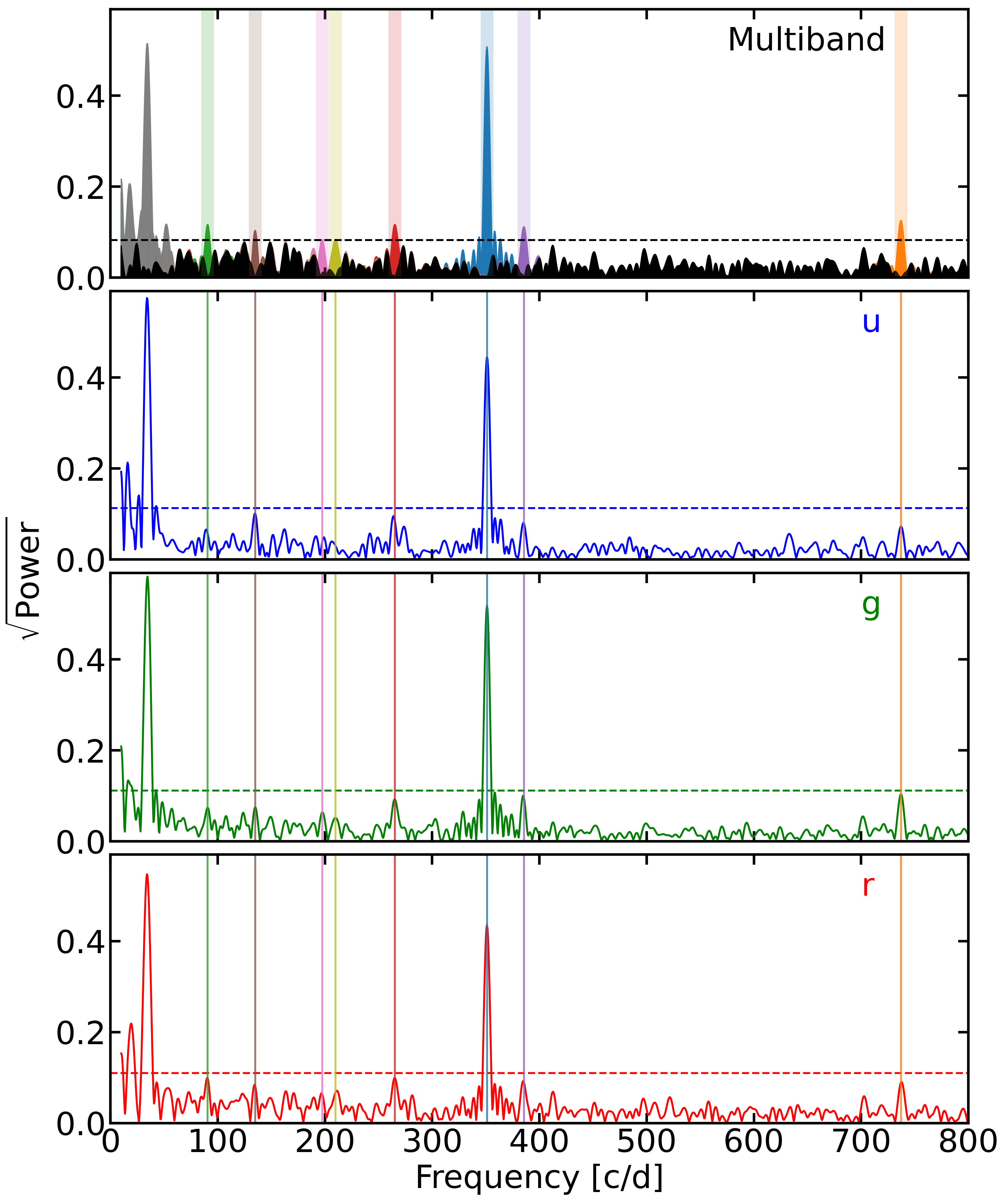}    
  \includegraphics[width=0.45\textwidth,angle=0]{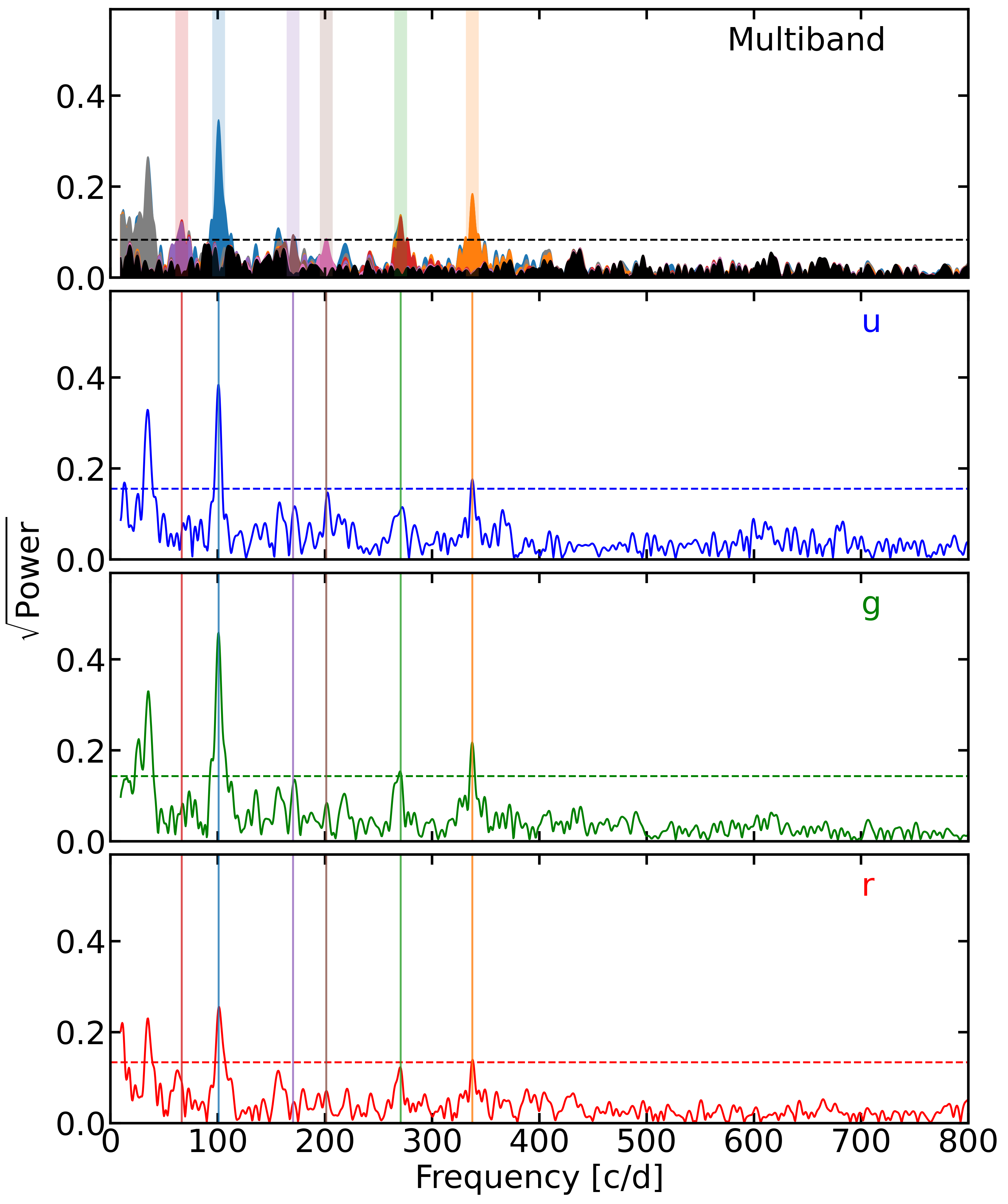} 
  \caption{Periodograms for the NTT/ULTRACAM data obtained in August 2022 (left panels) and September 2023 (right panels). The top panels show the multiband periodograms, while individual periodograms for each band ($u'$, $g'$, $r'$ from top to bottom) are shown below for comparison. The dashed horizontal line in each panel indicates the detection threshold calculated as explained in the text. In the multiband periodograms, different peaks identified above the threshold are plotted in different colours and indicated by vertical shaded areas, unless associated with the orbital period and/or low-frequency noise (these are shown in grey). In most cases, the periods also correspond to a peak in the individual periodograms, as indicated by vertical lines marking their positions, although they do not always exceed the threshold in the individual data sets.}
    \label{multiband_fts}
    \end{center}
\end{figure*}

\subsection{Fourier analysis of ULTRACAM photometry}

Given the evidence for multiple periodicities in the NTT/ULTRACAM data of ASASSN--14dx, we conducted a detailed period analysis for both observed nights. As previously mentioned, the same dominant periodicity is not present in both data sets; therefore, the two nights were analysed independently.

We first calculated a multiband periodogram for each night following the algorithm of \citet{{2015ApJ...812...18V}} implemented with {\tt gatspy}\footnote{\href{https://www.astroml.org/gatspy/}https://www.astroml.org/gatspy/}. This is essentially a weighted sum of the individual periodograms, which allows the identification of frequencies present in all data sets. To calculate a detection threshold for significant periods, we employed a Monte Carlo method in which the measured fluxes (and their uncertainties) are shuffled while the time vector is kept unchanged. This removes real periodicities but preserves the scatter of the data. This was repeated 5,000 times, and each time the maximum of the periodogram was recorded. The detection threshold was set as the 99th percentile of the maxima. In other words, in each case, there is a 1 per cent chance that a period with the observed amplitude could be generated by random scatter alone. The highest period above the threshold was then subtracted from the data in the form of a fitted sinusoid with that period, and the multiband periodogram and threshold were recalculated. This pre-whitening process was repeated until no new periods were found above the threshold amplitude. The resulting periodograms are shown in Fig.~\ref{multiband_fts}. For the first night of observations, we identified eight frequencies above the threshold, whereas six periodicities were found on the second night.

\begin{table*}
  \begin{tabular}{ccccccccc}
    \hline
 \multicolumn{2}{c}{Period (s)} & \multicolumn{2}{c}{Frequency (c/d)} & \multicolumn{2}{c}{ID} & \multicolumn{3}{c}{Amplitude (mmag)} \\
  2022 & 2023  & 2022 & 2023 &  2022 & 2023 & $u'$ & $g'$ & $r'$ \\
    \hline
$117.2\pm0.06$ & &  $737.06\pm0.40$ &    & $f_2$ &           & $0.86\pm0.34$ & $0.58\pm0.10$ & $<0.4$ \\
$224.03\pm0.25$ & &  $385.67\pm0.43$ &    & $f_2 - f_1$ &           & $0.90\pm0.34$ & $0.51\pm0.09$ & $0.23\pm0.06$ \\
$246.06\pm0.05$ & & $351.13\pm0.07$ &    & $f_1$ &      & $4.98\pm0.33$ & $2.85\pm0.09$ & $1.17\pm0.07$ \\
   & $256.04\pm0.20$ &  & $337.45\pm0.26$ & & $f_2$         & $2.59\pm0.46$ & $0.90\pm0.11$ & $0.46\pm0.10$ \\
   & $318.6\pm0.9$ &   & $271.1\pm0.7$ & & $f_3$          & $1.7\pm0.5$ & $0.60\pm0.11$ & $0.38\pm0.10$ \\
$326.5\pm0.8$ & &  $264.6\pm0.7$ &    & $f_3$ &               & $1.02\pm0.34$ & $0.35\pm0.20$ & $0.23\pm0.09$ \\
$418\pm5$ & &  $206.6\pm2.3$ &    & $f_2 - 2f_3$ &                 & $<1.7$ & $<0.5$ & $<0.3$ \\
   & $428\pm2$ &  & $202.0\pm1.0$ & & $2f_1$              & $1.9\pm0.6$ & $0.39\pm0.10$ & $0.2\pm0.1$ \\
$435\pm3$ & &  $198.4\pm1.2$ &    & $f_6$ &               & $<2.2$ & $<3$ & $<0.4$  \\
    & $506\pm2$ &   & $170\pm0.8$ & & $f_3 - f_1$            & $1.64\pm0.44$ & $0.53\pm0.10$ & $0.19\pm0.09$ \\
$640\pm2$ & &  $135.05\pm0.42$ &    & $f_4$ &                & $1.11\pm0.36$ & $0.38\pm0.09$ & $0.23\pm0.06$ \\
    & $856.0\pm1.2$ &  & $100.93\pm0.14$ & & $f_1$          & $5.2\pm0.5$ & $1.84\pm0.10$ & $0.80\pm0.24$ \\
$956\pm5$ & &  $90.4\pm0.5$ &    & $f_5$ &                 & $0.76\pm0.32$ & $0.31\pm0.17$ & $0.25\pm0.07$ \\
    & $1288\pm17$ &  & $67.0\pm0.9$ & & $f_3 - 2f_1$            & $1.51\pm0.46$ & $0.55\pm0.11$ & $0.34\pm0.11$ \\
\hline
  \end{tabular}
  \caption{The periods, frequencies, tentative identification, and amplitudes (or 3-$\sigma$ upper limits) of all the periodicities observed for ASASSN--14dx in the ULTRACAM data.}
  \label{periods}
\end{table*}

\begin{figure}
  \begin{center}
  \vspace{0mm}   
  \hspace{-10mm}  
  \includegraphics[width=0.52\textwidth,angle=0]{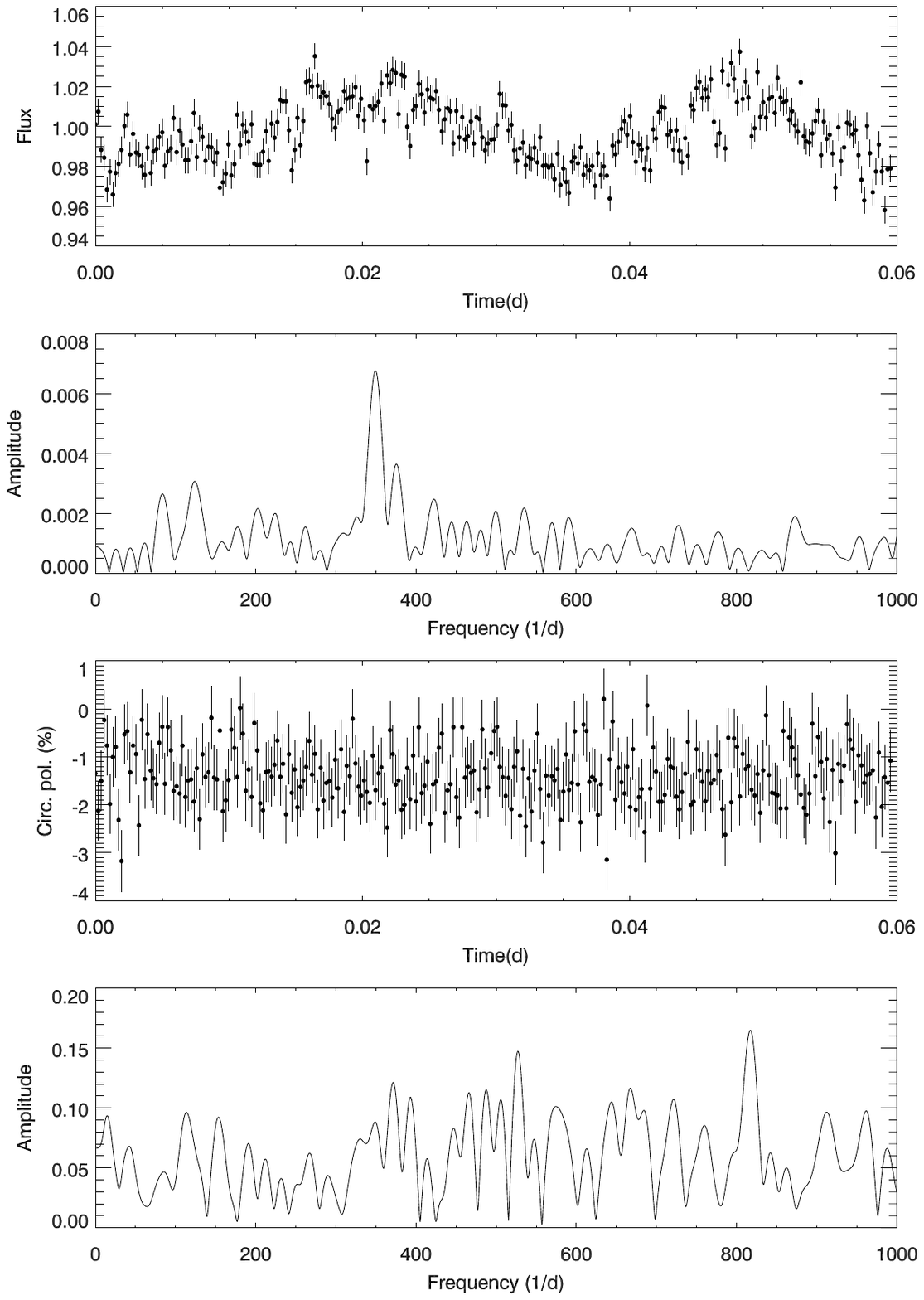}    
  \caption{High time-resolution NOT circular polarimetry of ASASSN–-14dx in a sky contrast filter. From top to bottom: the light curve covering one orbital period, the Lomb-Scargle amplitude spectrum of the pre-whitened light curve, the circular polarisation curve, and its amplitude spectrum. The zero calibration of the circular polarisation is not accurate in high-time-resolution mode, resulting in a $-1.3$\% shift in the overall level of circular polarisation.}
    \label{figurecpol}
    \end{center}
\end{figure}

Once a set of significant periodicities was identified for each data set, we performed a Fourier fit of the data to better constrain the periods and amplitudes. This was done by first subtracting a fit that included only the low-frequency components (shown in grey in Fig.~\ref{multiband_fts}) to remove orbital signatures, and then fitting the remaining periodicities. We simultaneously fitted the data from all bands with the same periods, but with individual amplitudes and phases for each filter. An example Fourier fit is shown in Fig.~\ref{fourier_fit}. Uncertainties were determined via bootstrapping, where we re-sampled the data, allowing for repetitions (essentially giving variable weight to measurements), and repeated the fit a thousand times. The obtained periods and amplitudes are reported in Table~\ref{periods}. As indicated in the table, preliminary analysis suggests that up to six independent frequencies were observed on the first night, and three on the second night, with the remaining peaks attributed to harmonics and linear combinations. No frequencies appear to be common between the two nights.

The presence of multiple frequencies, rather than a single frequency with harmonics that could be attributed to the white dwarf spin, suggests that ASASSN--14dx harbours a pulsating white dwarf. The range of periods detected is consistent with those observed for GW~Lib stars \citep{2021FrASS...8..184S}, named after the prototype accreting pulsating white dwarf \citep{1998IAUS..185..321W}.

\subsection{Circular polarimetry}

\begin{figure*}
  \begin{center}
\hspace{0mm}  
\vspace{0mm}
  \includegraphics[width=1.05\textwidth]{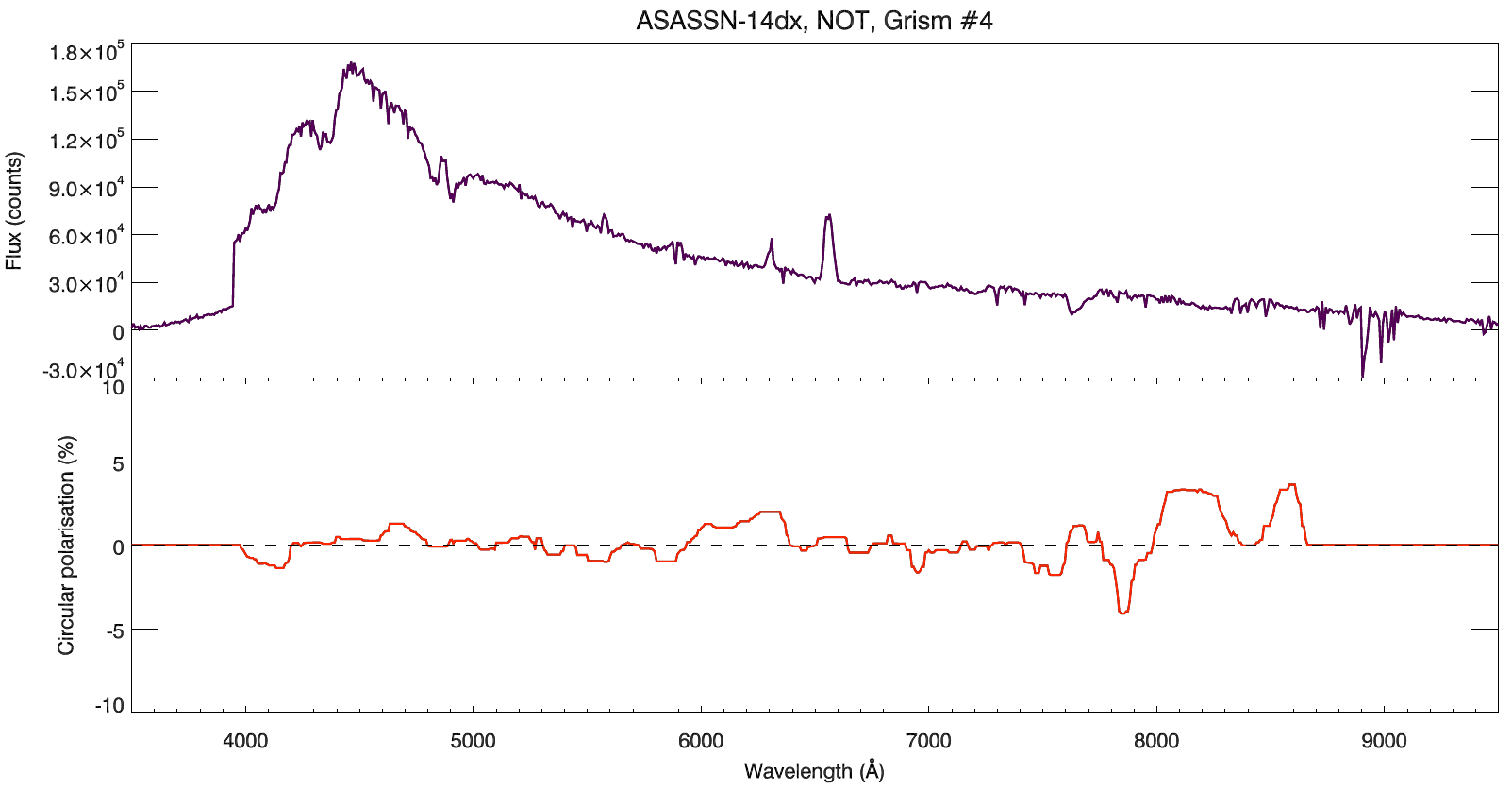}    
\vspace{-5mm}
  \caption{Circular spectropolarimetry of ASASSN--14dx from NOT. The polarisation spectrum (bottom) was median filtered with a 15 pixel window. There is no sign of circular polarisation.}
    \label{figurespecpol}
    \end{center}
\end{figure*}

Since both the NOT (31 Aug 2021) and NTT (22 Aug 2022) photometry appeared to indicate a single period, with no sidebands or harmonics, the original working hypothesis was that the 4-min period, observed in all other photometric datasets, apart from the first (26 Aug 2021) and last (4 Sep 2023) ones, likely represented the spin period of a magnetic white dwarf. This was particularly plausible given that the initial detection of the 14-min period was based on only a couple of cycles (26 Aug 2021). To investigate this possibility, we obtained time-series circular polarimetry and circular spectropolarimetry from the NOT (Figs.~\ref{figurecpol} and \ref{figurespecpol}).

The single circular polarimetric spectrum covered the wavelength range from 4000--8600\;\AA\ (Fig.~\ref{figurespecpol}). There is no indication for any circular polarisation in the continuum. As the spectropolarimetry covered only one epoch, and was constructed from four sub-exposures, we
also obtained fast time-series imaging circular polarimetry to properly sample the 4-min
cycle (Fig.~\ref{figurecpol}). These observations lasted for 86\;min, thus covering a full orbital period of the system. The data, taken in wide sky contrast filter (OG570) covering essentially the $g$+$r$ bands, show the orbital modulation and the 4-min period in photometry. However, as in the case of spectropolarimetry, there is no indication of the 4-min period in the circular polarisation.
Curiously, the photometry derived from the circular polarisation observations carried out on 27 and 28 November 2022 are in anti-phase with respect to the maxima and minima phases. The effect is real and visible in Fig.~\ref{figurephotom}. The circular polarisation mean level is offset by $\approx-1.3$\% due to the lack of a proper zero-level correction in the fast, fixed optical path, circular polarimetry mode utilised here.

\subsection{Optical spectroscopy}
In addition to the photometry and polarimetry, we obtained optical spectra from the SAAO 1.9-m telescope with the SpUpNIC  spectrograph \citep{spupnic2019} in 
March 2023. A total of four spectra were taken with grating \#6 (1.34\;\AA\;pix$^{-1}$ dispersion), yielding a spectral
resolution of $R \approx 1000$. The spectra were bias-subtracted and flatfielded using IRAF routines, extracted using a python implementation of the algorithm described by \citet{horne1986}, and wavelength calibrated using adjacent lamp spectra. In the absence of accurate orbital phase information, the four spectra were checked against mutual radial velocity shifts, and as none were found, they were combined into one. The average spectrum is dominated
by the blue continuum emission from the white dwarf, together with very broad Balmer absorption lines. There is also a clear indication of the presence of an accretion disc, as the Balmer line cores are filled with strong double-peaked emission lines. The observed features are consistent with those reported by \citet{2014ATel.6624....1K} and \citet{2016AJ....152..226T}. 

Subsequently, ASASSN--14dx was also observed as part of the 300 pc survey of CVs (Pala et al., in prep.) with the VLT and the X-Shooter spectrograph \citep{vernetetal11-1}. Two spectra were obtained using the 1.3-arcsec slit of the UVB arm ($R = 4100$) and the 1.2-arcsec slit of the VIS arm ($R = 6500$), with exposure times of 271 and 260\,s, respectively. The seeing, measured from the spectral traces, was approximately 1.5\;arcsec (UVB) and 1.1\;arcsec (VIS). The conditions were not photometric.

\begin{figure*}
  \begin{center}
  \vspace{0mm}   
  \hspace{0mm}  
  \includegraphics[width=1.05\textwidth,angle=0]{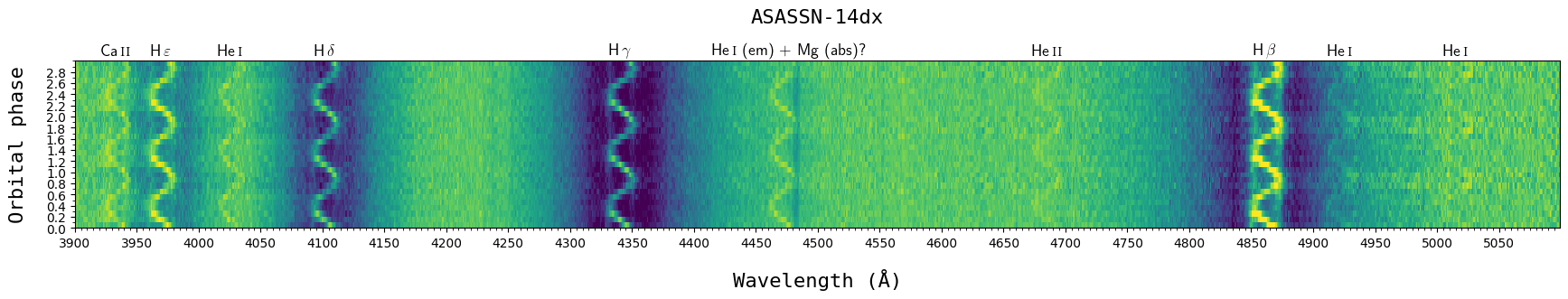}    
  \includegraphics[width=1.05\textwidth,angle=0]{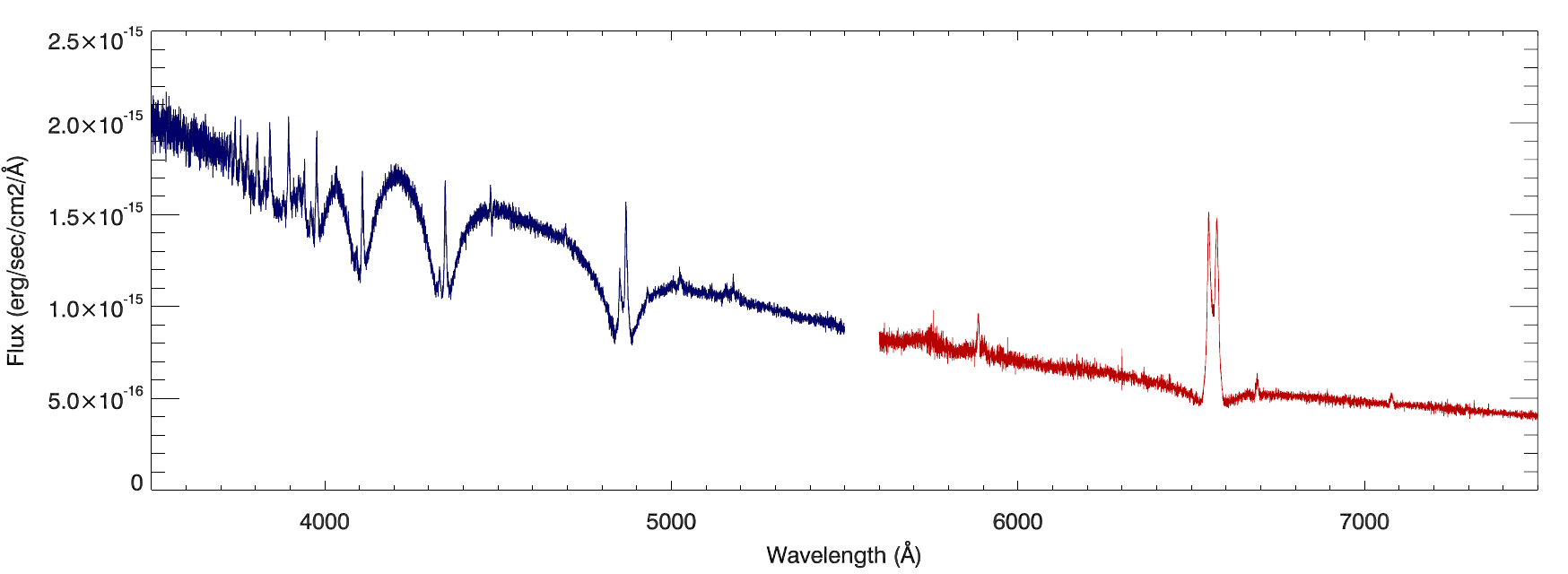} 
  \caption{The NOT trailed spectra binned into 10 phase bins and plotted over three orbital periods for clarity (top). The single epoch X-Shooter UVB+VIS band spectrum (bottom)}
    \label{figureNOTtrailed}
    \end{center}
\end{figure*}

As both the SAAO and X-Shooter spectra only covered a partial orbital period, or a single epoch in the case of X-Shooter, we obtained a full orbital period of phase-resolved spectra using the NOT in September 2024. The NOT spectra were taken using ALFOSC, equipped with grism \#18, which covers the 3500--5300\;\AA\ wavelength range with a resolution of $R=1000$. Both the NOT and X-Shooter spectra (Fig.~\ref{figureNOTtrailed}) clearly demonstrate that the asymmetry in the emission lines increases strongly towards the higher order Balmer lines, with \Ha\ showing an almost symmetric disc profile, as also seen in the SAAO spectra. Furthermore, the NOT spectra (Fig.~\ref{figureNOTtrailed}) reveal a sinusoidally modulated component in several emission lines. This component is much hotter 
than the accretion disc (judging by the relative Balmer decrement) and likely originates from the hot spot in the disc, where the accretion stream hits the disc's outer rim. A sinusoidal fit to the modulation from the hot spot reveals $K_\mathrm{spot} = 574.5 \pm 3.6$\;\kms\ (using the \hel{i}{4472} line), to which we will return later. The period was fixed to the orbital period when fitting for the radial velocity amplitude, although we note that we do not have sufficient time coverage that would allow us to study any subtle difference in the period.

\begin{figure}
  \begin{center}
  \includegraphics[width=0.5\textwidth,angle=0]{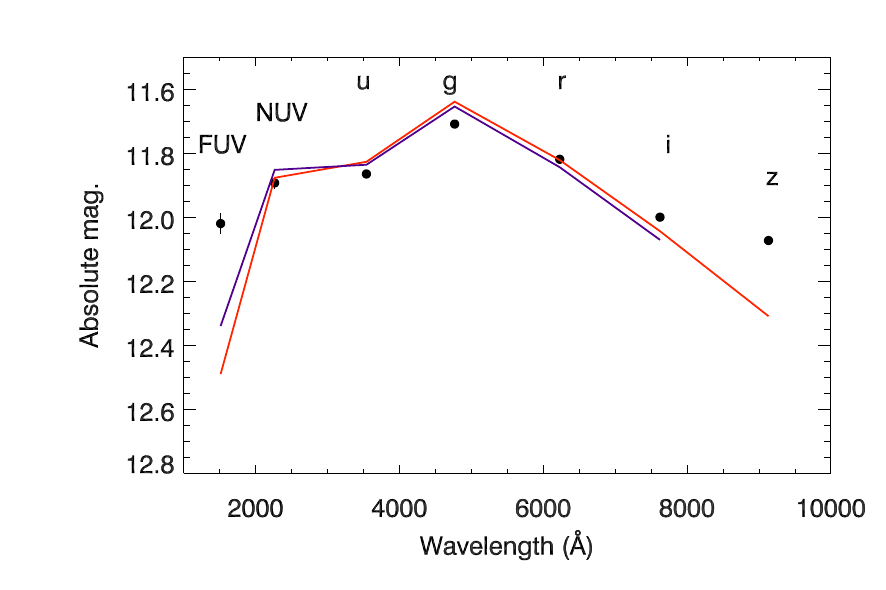}    
  \caption{The pre-outburst SED of ASASSN--14dx, along with the DA white dwarf model fits. The red fit uses the full FUV--SDSS $z$ range, whilst the magenta fit excludes the SDSS $z$-band, which appears to show evidence of NIR excess. Note the excess in the FUV band.}
    \label{figuresedfit}
    \end{center}
\end{figure}

\section{Modelling ASASSN--14dx}

In this section, we present the results of our modelling of the binary system. Specifically,
we performed both spectral and spectral energy distribution (SED) fitting, combined with fitting the \Ha\ emission line profile using an optically thin Keplerian disc model. We then attempt to integrate the fitting results into a unified model for the system.

\subsection{SED fitting}

We retrieved the SED data from Vizier \citep{2000A&AS..143...23O}, which includes data points spanning from the far-ultraviolet (\textit{GALEX} FUV) to the near-infrared (NIR; WISE W2), covering wavelengths from 1500\;\AA\ to
12\;$\mu$m. The photometric measurements show significant scatter and appear bimodal in flux, likely due to measurements being obtained both pre and post the outburst of 2014.  

To obtain the best possible constraints on the white dwarf properties, we restricted 
our SED modelling to data taken prior to the June 2014 outburst, as the long-term Zwicky Transient Facility (ZTF) light curve indicates that even ten years after the outburst, the source has not fully returned to its quiescent, pre-outburst state. This left us with the \textit{GALEX} FUV and NUV points, along with the Sloan Digital Sky Survey (SDSS) $ugriz$ magnitudes. The WISE IR points showed a noticeable IR excess, and where therefore excluded from the white dwarf model SED fit.

We used the \citet{2011ApJ...730..128T}\footnote{\href{https://www.astro.umontreal.ca/~bergeron/CoolingModels/}{https://www.astro.umontreal.ca/~bergeron/CoolingModels/}} pure-hydrogen atmosphere white dwarf models, which are tabulated in absolute magnitudes, as our source has a known distance. The models were interpolated from a grid of 305 DA models during the fitting. However, given the surface gravity ($\log g$) constraints implied by the spectral fitting (discussed below), we restricted the model grid to $\log g=8.75$, corresponding roughly to a white dwarf mass of 1.1\;$M_{\sun}$ for the temperature range in question. Initially, we fitted the \textit{GALEX} FUV, NUV, and SDSS $ugriz$ bands, which resulted in an effective temperature of $T_\mathrm{eff} = 13143 \pm 43$\;K (Fig.~\ref{figuresedfit}, red line). However, as there appears to be NIR excess present already in the $z$ band, we also fitted the SED excluding it (magenta line in Fig.~\ref{figuresedfit}). The best fit in this case yields $T_\mathrm{eff} = 13397 \pm 42$\;K. In both cases, additional flux is required to match the observed magnitudes. The required offsets are $-1.07 \pm 0.01$ and
$-1.02 \pm 0.01$\;mag for the FUV--SDSS $z$ and FUV--SDSS $i$ SED ranges, respectively. 
The likely explanation is that, even before the outburst, emission from the system was dominated by the accretion disc.
The contribution from any donor star appears non-existent. There are no spectral features, such as K I doublets, visible in the NIR arm of the X-shooter spectrum that could be attributed to a donor star.
The above mentioned $z$-band excess in SED is therefore either accretion disc related or due to intrinsic variability of the system. In fact, \citet{2023MNRAS.523.6114N} detect similar NIR excess in BW Scl, which is a very similar system to ASASSN--14dx. They suggest that at least part of it is due to the low temperature emission from optically thin outer disc. Currently we also cannot entirely rule out a very low field (B$<$ 1-5 MG) white dwarf that could contribute to NIR flux via cyclotron emission.

If we adopt the DA white dwarf model based on the FUV--SDSS $i$ energy range, we note that the \textit{GALEX} FUV shows an excess of approximately 0.3\;mag, while there is no excess in the \textit{GALEX} NUV band. One approch to explain this would be via a small hot spot (with a fractional projected area of the order of 10$^{-4}$) on the white dwarf surface, emitting very soft X-rays with a blackbody temperature of $kT\sim 30$\;eV, as observed in polar CVs \citep{2008xru..confE..72R}. This is estimated by convolving two blackbody spectra corresponding to the $T_\mathrm{eff} = 13397K$ and $T_\mathrm{BB} = 350000K$ with the \textit{GALEX} FUV and NUV response curves and converting to magnitudes. A perhaps more likely explanation, and the one we favour, for such excess is the possible presence of immensely strong CIV 1550\AA\ emission line, as seen in SDSS J123813.73–033933.0 \citep{2019MNRAS.483.1080P} and especially in GW Lib \citep{2002ApJ...575L..79S}. The \textit{GALEX} FUV band has the maximum efficiency exactly around the CIV 1550\AA\ line and the FUV magnitude would certainly be affected by the presence of such line.

In the case of a weakly magnetic white dwarf, the inner accretion flow could be channelled along the field lines onto the magnetic pole(s) of the white dwarf, leading to such emission. Hot spots on the white dwarf have been observed in a very similar system SDSS J123813.73–033933.0, and were associated to the spiral shocks in the accretion disc \citep{2019MNRAS.483.1080P} rather than a magnetic field. We should note though, that there is no persistent period detected in our photometry that would be expected from such hot spot(s) on the white dwarf surface, strongly suggesting that the FUV excess is a result of strong CIV 1550\AA\ emission.

\begin{figure*}
  \begin{center}
  \vspace{0mm}   
  \hspace{5mm}  
  \includegraphics[width=1.0\textwidth,angle=0]{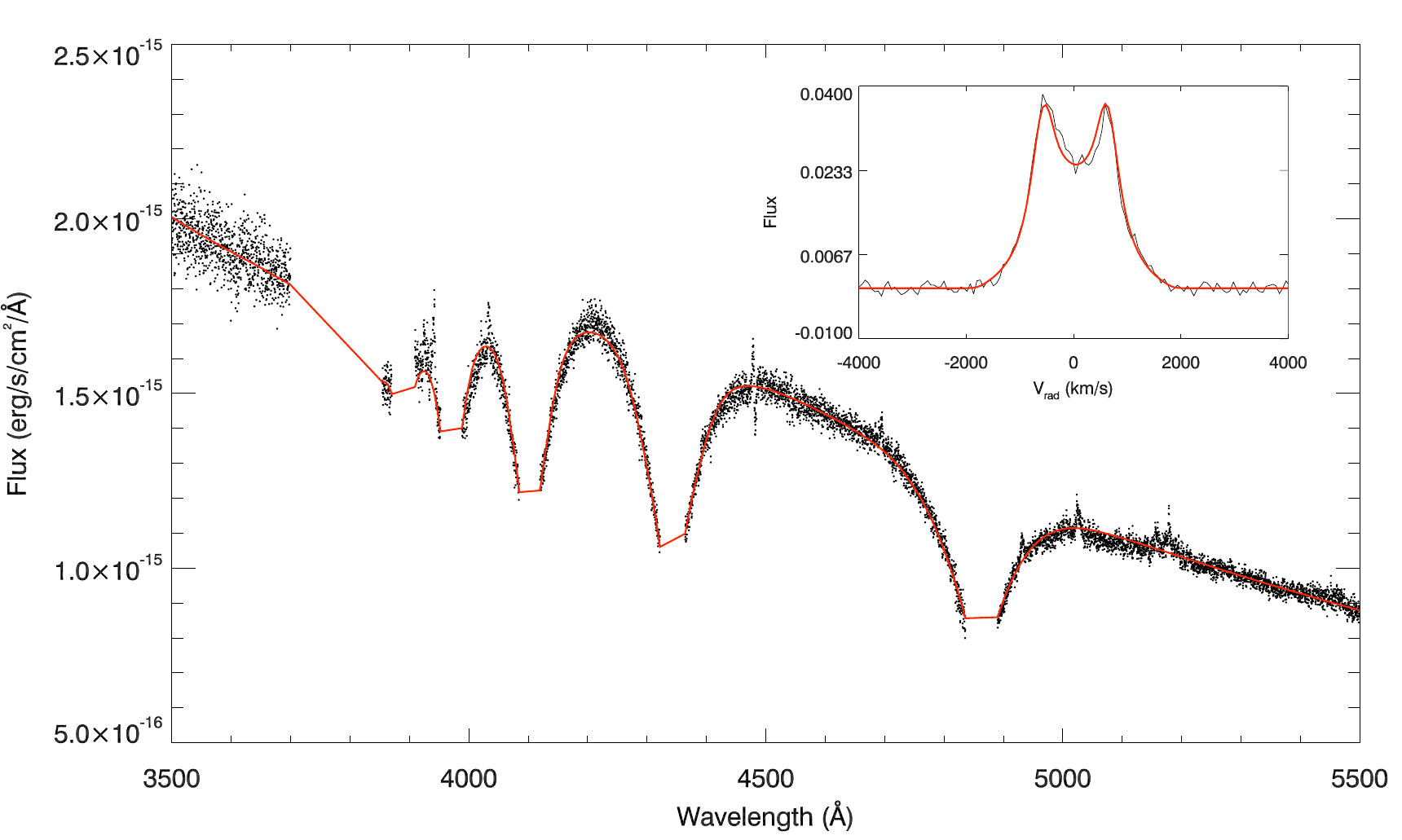}    
  \caption{The combined Koester DA white dwarf model and accretion disc model fitted to the mean, flux-calibrated (NOTE, non-photometric conditions!) X-Shooter UVB spectrum is shown in the main plot. The \Ha\ profile from the SAAO 1.9-m, after subtracting the continuum plus absorption line fit, averaged over 40 min, is shown alongside the best-fit Keplerian disc model (inset). The emission lines redward of H\textbeta\ are likely \Line{O}{iii}{4932}, \Line{N}{ii}{5026}, \Line{Fe}{vii}{5158}, and \Line{N}{ii}{5180}.}
    \label{figurefits}
    \end{center}
\end{figure*}

\subsection{Spectral and \Ha\ emission line profile fitting}

To fit the overall optical spectrum, we used a model consisting of three components. The spectrum was fitted using the Koester DA white dwarf models \citep{2010MmSAI..81..921K} combined with an accretion disc model \citep{1997MNRAS.289..388G}. The disc model does not include an emission line profile model for Keplerian discs. Therefore, the \Ha\ profile was fitted separately using a Keplerian accretion disc kinematic model, where the line emission is represented as a histogram of radial velocities from a Keplerian disc with a specified radial emissivity profile. 

The free parameters for this kinematic model are the system inclination, white dwarf mass, radial emissivity power index ($\alpha$), disc inner radius ($r_\mathrm{in}$), disc outer radius ($r_\mathrm{out}$), and the mean radial velocity ($\gamma$). 
The free parameters for the Koester DA white dwarf model are the white dwarf $T_\mathrm{eff}$ and $\log g$, along with the white dwarf radius for scaling its flux density (as the distance is fixed by \textit{Gaia}). The free parameters for the accretion disc isothermal slab model are the gas pressure ($P_\mathrm{gas}$), gas temperature ($T_\mathrm{gas}$), slab height ($z$) and inclination. 

Although the emission lines from the accretion disc model provide valuable information about the disc temperature via the Balmer decrement, we had to exclude the line cores from the spectral fitting. This is because the higher-order Balmer lines are skewed and highly variable over the orbital period due to the movement of the hot spot (see Fig.~\ref{figureNOTtrailed}), and the X-Shooter data were obtained during a single epoch. The gaseous slab model used here cannot generate emission line profiles for non-uniform disc surface brightness (or temperature) distributions. As a result, all spectral regions showing evidence of line emission were excluded from the overall spectral modelling. This means that the accretion disc model contributes only as a continuum component. For a similar reason, we used the SAAO spectra for the \Ha\ modelling. The SAAO spectra cover at least half of the orbital cycle, and \Ha\ is least affected by the hot spot. Furthermore, the SAAO \Ha\ emission line profile appears completely symmetrical. 

The fit was performed using the Markov Chain Monte Carlo (MCMC) method with an affine-invariant stretch algorithm \citep{GW2010} for sampling, employing 100 random walkers. A bespoke implementation of the algorithm was used. Since the individual spectra were obtained with a narrow slit to enable accurate radial velocity measurements, the absolute flux calibration is somewhat unreliable due to slit losses. Moreover, the conditions during the X-shooter observations were non-photometric adding further to the uncertainty of the true flux level.

The X-Shooter UVB spectrum, along with the best fit, is shown in Fig.~\ref{figurefits}. The inset displays the Keplerian accretion disc fit to the SAAO \Ha\ emission line profile, obtained simultaneously with the overall spectral fit using the Koester DA white dwarf model combined with an isothermal gaseous slab accretion disc model (main panel). The corner plot of the MCMC modelling is shown in Fig.~\ref{figuremcmc}, and the best-fit parameters, along with their error limits derived from the MCMC posterior marginal distributions, are listed in Table~\ref{table2}.

\begin{table}
  \begin{tabular}{lllll}
    \toprule\noalign{\smallskip}
Parameter           & Median & $1\sigma$ & 99.9\%> & 99.9\%< \\
\noalign{\smallskip}    \midrule\noalign{\smallskip}
 Inclination (deg) & 49.0 &   --4 +5 &  35.0 & 63.0 \\
 WD mass ($\mathrm{M}_\odot$) & 1.00 & --0.15 +0.17  & 0.67&  1.39\\
 $\alpha$ &  --1.16  &   --0.06 +0.05  & --1.33 & --1.02\\
 Disc r$_\mathrm{in}\;(a)$  &  0.06  & --0.01 +0.01  & 0.04 & 0.08\\
 Disc r$_\mathrm{out}\;(a)$ &   0.53  &  --0.04 +0.04   &  0.39 & 0.60\\
 $\gamma$ (\kms) & --34.5  &  --1.8 +1.8  & --41 & --28\\
 $T_\mathrm{eff}$ (K) & 16140  & --68 +73 & 15910 & 16357 \\
 $\log g$ (cgs) & 8.767 & --0.017 +0.015 & 8.717 & 8.836\\
 WD radius (km) & 5448 & --30 +29 & 5356 & 5543\\
 $P_\mathrm{gas}$ (dyn\;cm$^{-2}$) & 196  & --42 +46 & 100 & 468\\
 $T_\mathrm{gas}$ (K) & 6093  & --15 +16  & 6048 & 6144\\
 $\log z$ (cgs) &8.12  &  --0.12 +0.14  & 7.80 & 8.44 \\
\noalign{\smallskip}\bottomrule
  \end{tabular}
  \caption{The MCMC results. $a$ is the binary separation. {\bf Note} that the listed WD mass only comes from the emission line profile fitting (together with the inclination). Furthermore, the quoted WD radius serves ONLY as a scaling factor for the model to match the (non-photometric) flux level.
  $T_\mathrm{eff}$ and $\log g$ are constrained by the spectral fit itself.
  }
  \label{table2}
\end{table}

\subsection{Gravitational redshift}

The NOT and X-Shooter spectra show evidence of a narrow absorption line of \Line{Mg}{ii}{4481}, formed in the white dwarf atmosphere. This line serves as a tracer for the gravitational redshift produced by the white dwarf's potential well and, consequently, can be used to estimate the white dwarf mass. The NOT spectra cover the full orbit of ASASSN--14dx, allowing the fitting of trailed line profiles to account for the modulation caused by the hot spot, which can be used to determine the systemic velocity, $\gamma$. Using \Line{He}{i}{4471}, we obtained $\gamma=-30.3\pm 2.7$\;\kms.
Summing all the NOT spectra yields an average \Line{Mg}{ii}{4481} profile, whose central wavelength reflects only the system's redshift, $\gamma$, combined with the gravitational redshift. We measure a central redshift of $76.6 \pm 6.9$\;\kms. Subtracting the systemic redshift gives a gravitational redshift of $106.9 \pm 7.4$\;\kms. This corresponds to an independent  measurement of the white dwarf surface gravity and mass of $\log g=8.84\pm0.05$ and $1.12\pm0.02\;\mathrm{M}_{\odot}$. These values are consistent with our spectral fitting, which yielded $\log g=8.77\pm0.02$.

\section{Discussion}

Our photometric, polarimetric, and spectroscopic observations suggest that ASASSN--14dx is likely a short orbital period CV harbouring a pulsating DA white dwarf. The photometry reveals two distinct dominant periods of 4 and 14\;min, which fall within the typical range for pulsating CVs \citep{2021FrASS...8..184S}. Based on fits using Koester DA white dwarf models, the spectroscopy suggests the presence of a relatively massive white dwarf.

\subsection{Stellar and orbital parameters}

The combined MCMC analysis of the overall spectral fitting and the \Ha\ line profile fitting implies a white dwarf effective temperature of $T_\mathrm{eff} = 16140^{+73}_{-68}$\;K and $\log g = 8.77_{-0.02}^{+0.02}$. Comparing these values with the Montreal database of white dwarf evolutionary models \citep{2017ASPC..509....3D, 2020ApJ...901...93B} suggests a white dwarf mass of $1.09 \pm 0.01\;\mathrm{M}_{\odot}$ and a radius of $0.0071 R_{\odot}$ (4940\;km) for the resulting $\log g$ value. Using La Plata white dwarf models \citep{2016ApJ...823..158C}\footnote{\href{http://evolgroup.fcaglp.unlp.edu.ar/TRACKS/tracks.html}{http://evolgroup.fcaglp.unlp.edu.ar/TRACKS/tracks.html}} yields a mass of $1.06 \pm 0.01\;\mathrm{M}_{\odot}$ and a similar radius of $0.0071R_{\odot}$.
We note that our best-fit white dwarf radius, derived from the flux scaling of our model, is $0.00783 R_{\odot}$, with a 99.9\% lower limit of $0.00770 R_{\odot}$. However, the flux level of the X-shooter spectrum is not reliably calibrated due to slit losses and nonphotometric conditions. 

The effective temperature obtained from the spectral fitting disagrees with the SED fitting, which implied $T_\mathrm{eff} = 13400$\;K using a fixed $\log g=8.75$. However, it is important to note that the SED fitting did not include an accretion disc component due to the small number of data points. 
Furthermore, the SED data were obtained prior to the June 2014 outburst, and the source remains approximately 0.4\;mag brighter than its quiescent level even now, ten years after the outburst. 
Thus, it may not be appropriate to compare the two temperatures directly.

The orbital inclination of the binary is simultaneously fitted in both the spectral model fit (via the accretion disc model, which affects the flux scaling and the emergent continuum shape) and the line profile fitting. In the latter, the line profile width and shape are determined by the inclination and the white dwarf mass together. 

The moving hot spot in the NOT phase-resolved spectra (Fig.~\ref{figureNOTtrailed}), with a velocity amplitude of $K_\mathrm{spot} = 574.5 \pm 3.6$\;\kms, is best explained as the impact zone of the accretion stream on the outer edge of the disc. This velocity amplitude is consistent with such an explanation if we assume the fitted system parameters (Table~\ref{table2}) and a secondary mass of approximately $0.15\;\mathrm{M}_{\odot}$, based on the orbital period. Moreover, the coincidence of the $K_\mathrm{spot}$ velocity with the double peak
separation of the accretion disc emission line profile further strengthens this interpretation. 

\subsection{The nature of the photometric variability}

Apart from being somewhat overluminous, ASASSN--14dx 
appears to be a fairly typical CV that exhibits a pulsating white dwarf as its accretor. \citet{2021FrASS...8..184S} lists the current population of known CVs harbouring pulsating white dwarfs, comprising 18 systems with orbital periods clustering around the CV period minimum. ASASSN--14dx fits well within this group, although it has the highest white dwarf effective temperature among them. \citet{2021FrASS...8..184S} also lists the detected pulsation periods of these systems, which are very similar to the periods detected in ASASSN--14dx (Table~\ref{periods}).

In fact, the properties of ASASSN--14dx  remarkably resemble those of GW Lib \citep{2000BaltA...9..231V,2002ApJ...575L..79S}, both in terms of the detected pulsation periods and the white dwarf properties, although both the white dwarf effective temperature and mass are somewhat higher in the case of ASASSN--14dx. While the derived white dwarf temperature ($T_\mathrm{eff} = 16140$\;K) is well above the traditional range for the instability strip for non-accreting hydrogen-atmosphere white dwarf pulsators (ZZ Ceti stars), \citet{2006ApJ...643L.119A} demonstrated that an increased helium fraction from accretion and/or extreme surface gravity can extend the instability strip up to $T_\mathrm{eff} = 20000$\;K. This explains the population of GW Lib pulsators, to which ASASSN--14dx is a new addition.    

The change in the dominant mode observed for ASASSN--14dx is also consistent with a GW~Lib nature, with other systems also observed to show different dominant frequencies depending on the time of observation \citep[e.g.][]{2007ApJ...667..433M}. Changes in a short timescale, as the ones observed, are not necessarily indicative of changes to the stellar structure, which is expected to change at a much slower timescale. Instead, it is more likely that a different set of eigenmodes corresponding to the same stellar structure become excited, resulting in a different pulsation spectrum. In fact, while it has been observed that outburst can cause GW~Lib pulsations to disappear as the white dwarf is heated beyond the instability strip, the pulsations can come back with the same frequencies as observed pre-outburst, suggesting that outbursts do not singificantly affect the stellar interior \citep{2011ApJ...728L..33M}.

\subsection{Accretion rate and ASASSN--14dx as a (non)potential SNIa progenitor}

As ASASSN--14dx contains a high-mass, accreting  white dwarf, it is pertinent to consider the system's status as a potential SNIa candidate. \citet{2019PASJ...71...22I} classified ASASSN--14dx as a WZ Sge-class dwarf nova, primarily based on its discovery outburst properties and the 82.8-min orbital period. WZ Sge systems also have very long recurrence periods, typically of the order of decades. Currently, we know of only the 2014 discovery outburst for ASASSN--14dx, which is consistent with the WZ Sge classification. These systems also typically have very low accretion rates, which are thought to explain their long recurrence periods ($L_\mathrm{X} = 2.3 \times 10^{30}$\;erg\;s$^{-1}$ for WZ Sge itself in quiescence \cite{2004RMxAC..20..244M}, using the XMM-Newton and its 0.2--10\;keV band). The XMM-Newton 4XMM-DR14 catalogue \citep{2020A&A...641A.136W} includes two instances of ASASSN--14dx, both obtained in 2011, prior to the 2014 outburst. The stronger of these detections implies $F_\mathrm{X} = 5.7 \times 10^{-13}$\;erg\;s$^{-1}$\;cm$^{-2}$ in the 0.2--10\;keV band. Adopting the \textit{Gaia} distance, this translates to $L_\mathrm{X} = 4.5 \times 10^{29}$\;erg\;s$^{-1}$. This value is nearly an order of magnitude fainter than the WZ Sge value mentioned above and is among the faintest for dwarf novae, apart from GW Lib \citep{2010MNRAS.408.2298B}. Furthermore, assuming a typical accretion efficiency of 0.03\% for a white dwarf accretor, the measured X-ray luminosity corresponds to an accretion rate of $\dot{M} = 1.7 \times 10^{12}$\;g\;s$^{-1}$ (or $2.6 \times 10^{-14}\;\mathrm{M}_{\odot}$\;yr$^{-1}$).

Therefore, even if the white dwarf in ASASSN--14dx has a high mass, its current (and future) rate of growth is likely minimal. The white dwarf was either born massive or has accreted more mass than typical white dwarfs in CVs during its evolutionary history. There is not enough mass left in the secondary star to ever push the white dwarf over the Chandrasekhar limit. We thus conclude that ASASSN--14dx is not a viable SNIa candidate. The same applies to WZ Sge systems in general, unless their white dwarf mass is already extremely close to the Chandrasekhar limit.      

\section{Conclusions}

We have presented optical photometry, polarimetry and spectroscopy of ASASSN--14dx. Our data indicate that the system is a disc-accreting CV with a pulsating white dwarf accretor. There is no clear evidence for magnetism in the white dwarf based on our optical circular polarimetry. However, both the NIR/IR and the FUV excesses of the system could be explained with cyclotron emission from a low-field (less than a few MG) magnetic white dwarf. Such emission might also partially explain the overluminous nature of the system, but there is no evidence for cyclotron humps in the X-shooter NIR spectrum. The NIR excess can also be reconciled by emission from cool outer disc, but this would leave the FUV excess unexplained. The required combination of surface area and temperature to produce the observed FUV excess without any NUV excess appears to rule out equatorial hot regions. However, even if there is no conclusive evidence, we suspect that the FUV excess is most likely due to very strong CIV 1550\AA\ emission, that is centered on the FUV band and has been detected in ``sibling" systems GW Lib \citep{2002ApJ...575L..79S} and SDSS J123813.73–033933.0 \citep{2019MNRAS.483.1080P}.  

Our multi-epoch fast photometry reveals multiple pulsation periods, in particular at 4 and 14 min. Both are observed more than once, and one of them is always present. However, only one of them appears to be present at any given epoch. This is consistent with different sets of eigenfrequencies being observed at different epochs, as seen for other GW~Lib systems. We do not have a credible explanation for the 0.5 phase shift evident in the NOT photometry (Fig.~\ref{figurephotom}), unless the 4-min period results from the white dwarf spin and we would have witnessed a flip in accreting magnetic pole. However, we consider this scenario very unlikely.

Our optical spectral modelling, together with the measured gravitational redshift, suggests that the white dwarf in the system has a high mass, of the order of 1.1\;$\mathrm{M}_\odot$, while its spectroscopic effective temperature is $T_\mathrm{eff} = 16140$\;K, derived from combined spectral modelling using both a DA white dwarf and an accretion disc model. This temperature is higher than that of any accreting white dwarf pulsator system listed by \citet{2021FrASS...8..184S}, but consistent with theoretical expectations for GW~Lib-type pulsators, and GW Lib itself \citep{2016MNRAS.459.3929T}. Finally, NIR circular polarimetry could provide further insight into the possible magnetic nature of the white dwarf by confirming or ruling out the presence of cyclotron emission from a low-field magnetic white dwarf ($B \lesssim 10$\;MG).

\section*{Acknowledgements}

This research has made use of the VizieR catalogue access tool, CDS, Strasbourg, France. 

Based on observations made with the Nordic Optical Telescope, owned in collaboration by the University of Turku and Aarhus University, and operated jointly by Aarhus University, the University of Turku and the University of Oslo, representing Denmark, Finland and Norway, the University of Iceland and Stockholm University at the Observatorio del Roque de los Muchachos, La Palma, Spain, of the Instituto de Astrofisica de Canarias. Part of the data presented here were obtained with ALFOSC, which is provided by the Instituto de Astrofisíca de Andalucia (IAA) under a joint agreement with the University of Copenhagen and NOT. We would like to thank the NOT staff for carrying out some of the observations. 
Based on observations collected at the European Organisation for Astronomical Research in the Southern Hemisphere under ESO programmes: 109.23KC.001, 111.24HN.001 and 112.25Z0.001.

IP acknowledges support from a Royal Society University Research Fellowship (URF\textbackslash R1\textbackslash 231496). This research received funding from the European Research Council under the European Union’s Horizon 2020 research and innovation programme number 101020057 (JTW, BTG, SS). 
PR-G acknowledges support by the
Spanish Ministry of Science via the Plan de Generación de Conocimiento project
PID2021--124879NB--I00.

\section*{Data Availability}

The data will be shared on reasonable request to the corresponding author.

\vspace{4mm}

\bibliographystyle{mnras}

\appendix

\section{Fourier fits}

\begin{figure*}
  \begin{center}
  \vspace{0mm}   
  \hspace{0mm}  
  \includegraphics[width=0.49\textwidth,angle=0]{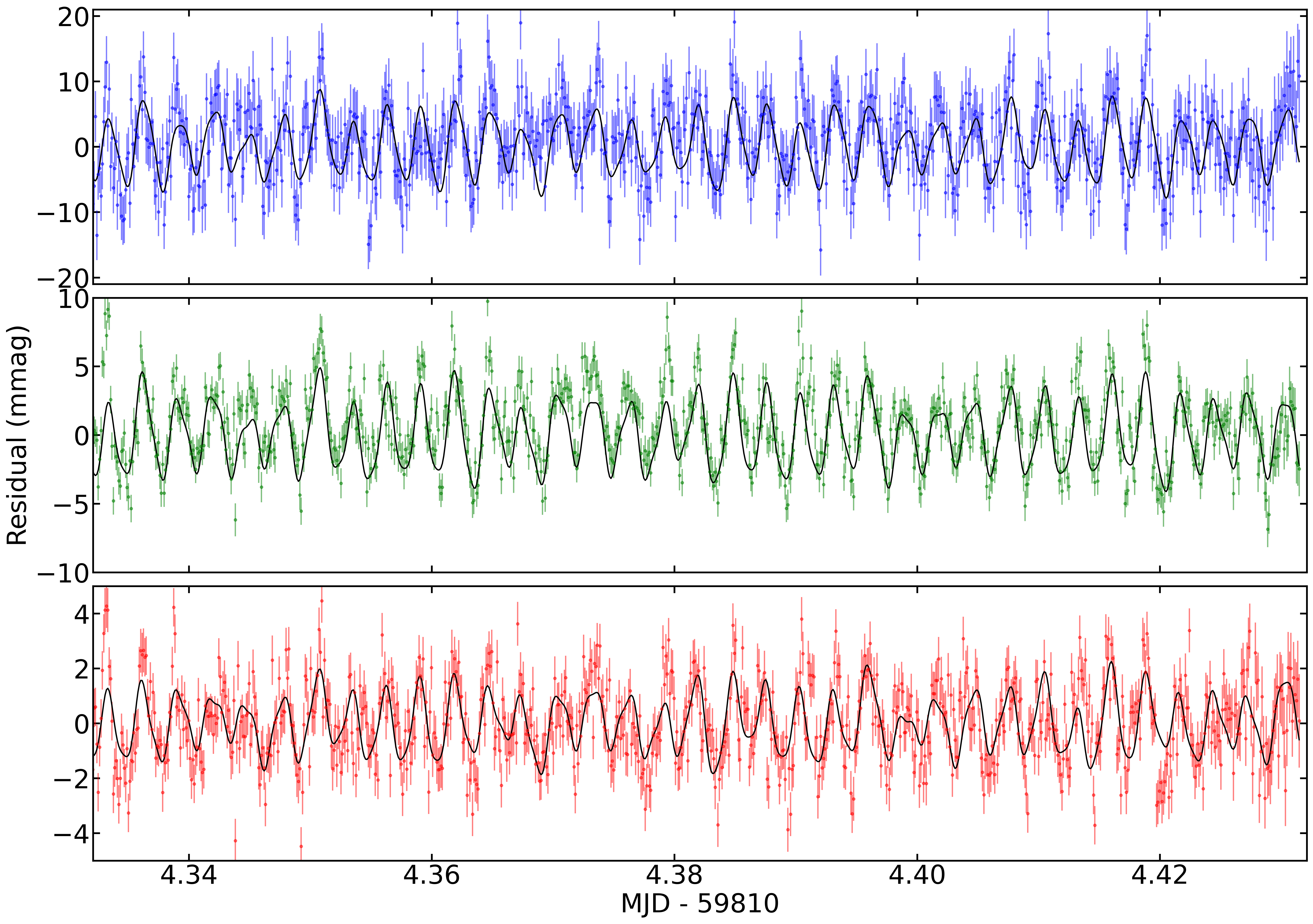}    
  \includegraphics[width=0.49\textwidth,angle=0]{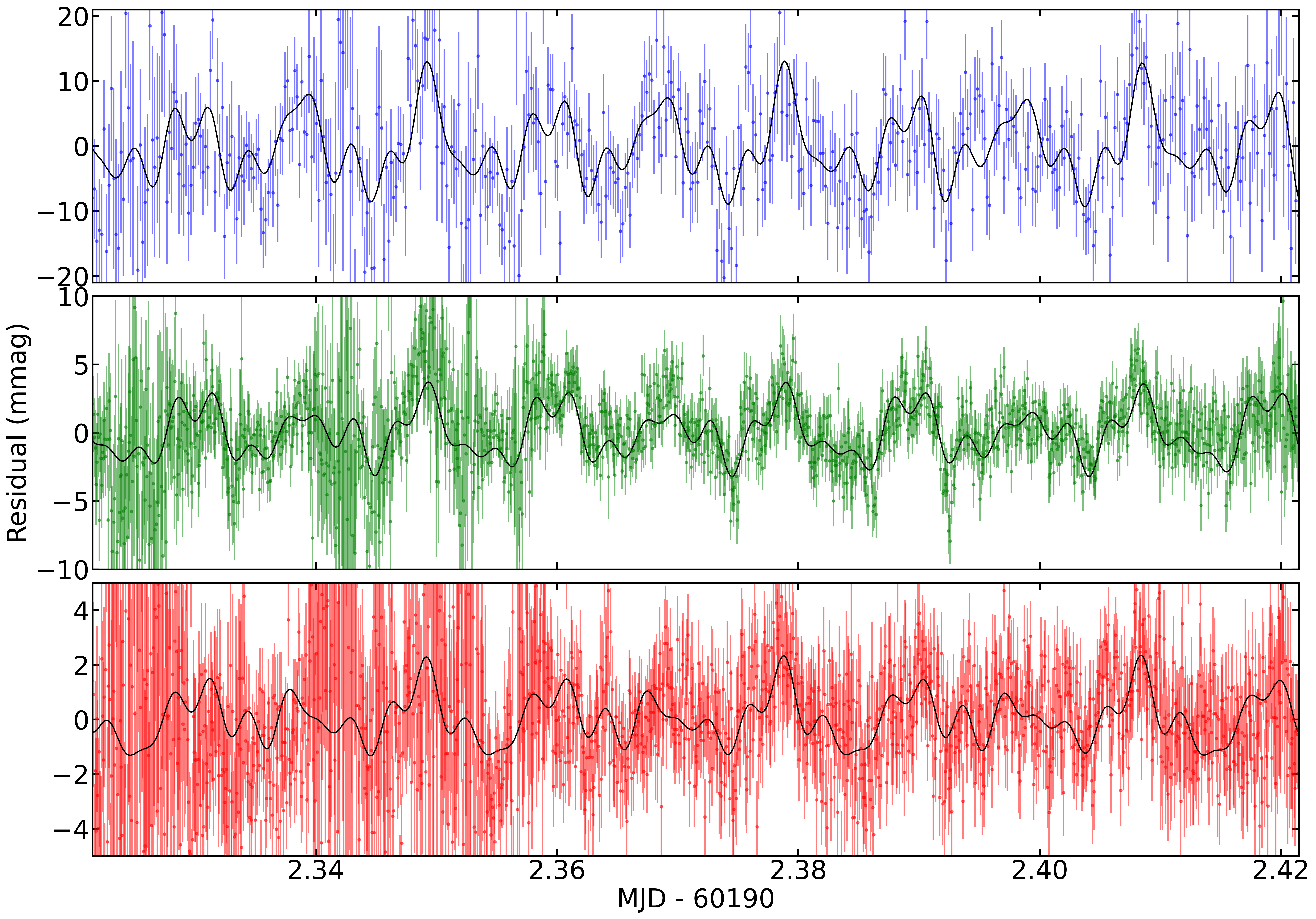} 
  \caption{Fourier fit for a section of the data taken in August 2022 (left) and September 2023 (right). The fit was performed to the residuals after low frequencies associated with the orbital period or noise were subtracted. The panels from top to bottom show the $u'$, $g'$, and $r'$ bands, with the best fit shown as a black line.}
    \label{fourier_fit}
    \end{center}
\end{figure*}

\section{MCMC results}

\begin{figure*}
  \begin{center}
  \vspace{0mm}   
  \hspace{-5mm}  
  \includegraphics[width=1.05\textwidth,angle=0]{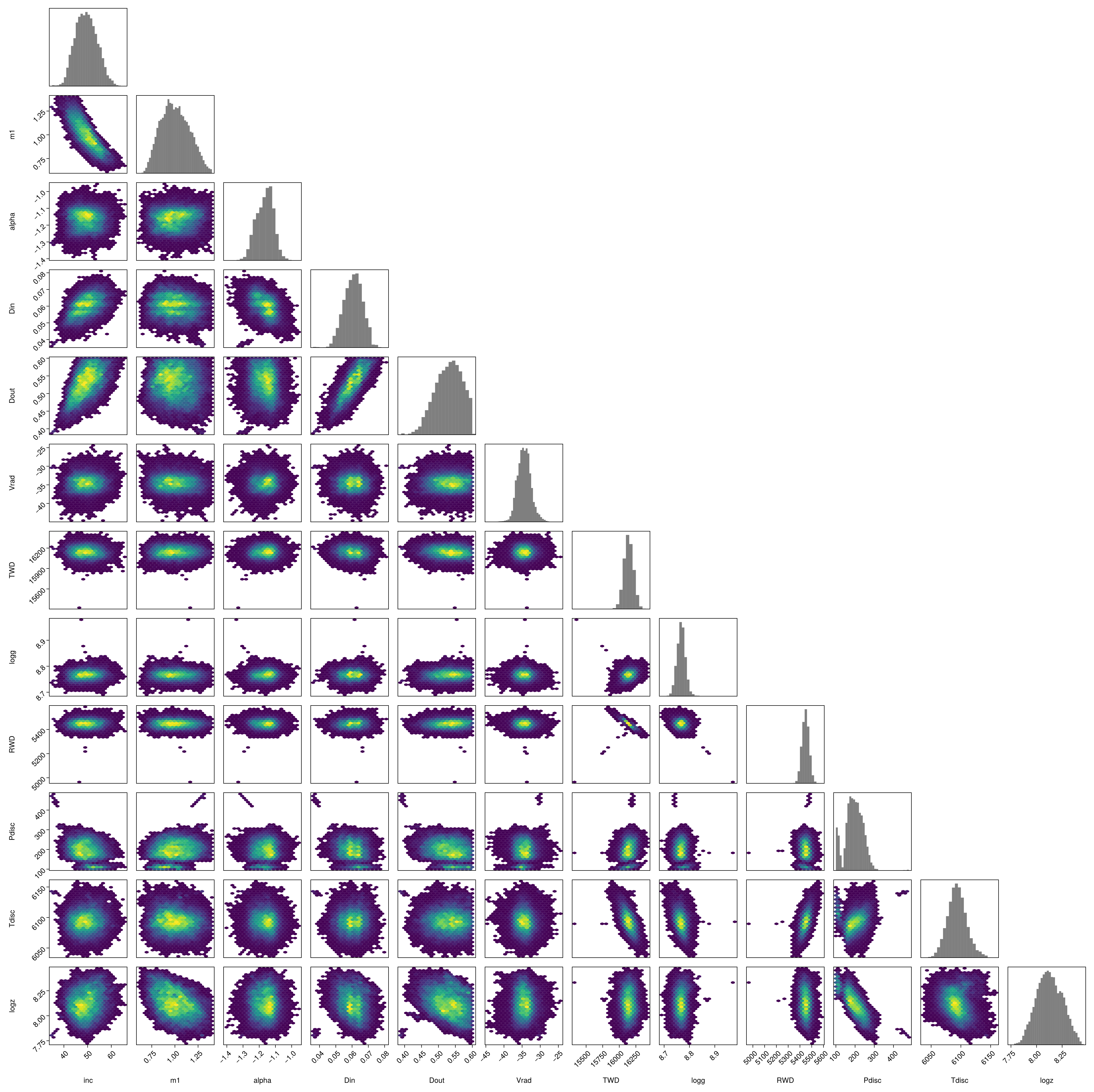}   
  \caption{
  The MCMC correlation and marginal probabilities plot of the \Ha\ emission line modelling and the white dwarf spectral fitting.}
    \label{figuremcmc}
    \end{center}
\end{figure*}

\bsp	
\label{lastpage}

\end{document}